\documentclass[aip,floatfix,amsmath,amssymb,reprint,superscriptaddress]{revtex4-2}

\usepackage{graphicx}
\usepackage{dcolumn}
\usepackage{bm}
\usepackage[colorlinks=True,allcolors=blue]{hyperref}
\usepackage{newtxtext,newtxmath}
\usepackage{microtype}
\usepackage{orcidlink}
\usepackage{lipsum}
\usepackage{amsmath,amssymb}
\usepackage{pdfpages}
\usepackage{sidecap}
\usepackage[detect-all]{siunitx}

\makeatletter 
\AtBeginDocument{\let\LS@rot\@undefined}
\makeatother
\makeatletter
\renewcommand\@make@capt@title[2]{%
 \@ifx@empty\float@link{\@firstofone}{\expandafter\href\expandafter{\float@link}}%
  {\textbf{#1}}\@caption@fignum@sep#2\quad
}%
\makeatother

\newcommand{\kB}{k_{\text{B}}}
\newcommand{\kT}{\kB T}

\newcommand{\Dxi}{\Delta x_{\text{i}}}

\newcommand{\vg}{v}

\newcommand{\etaltxt}{\latin{et al.}} 
\newcommand{\leibnizd}{\mathrm{d}}
\newcommand{\dt}{\leibnizd{t}}
\newcommand{\ddt}[1]{\leibnizd{#1}/\dt}

\newcommand{\Pe}{\mathrm{Pe}}
\newcommand{\const}{\mathrm{constant}}

\newcommand{\Eq}[1]{Eq.~\ref{#1}}
\newcommand{\Eqs}[1]{Eqs.~\ref{#1}}
\newcommand{\Fig}[1]{Fig.~\ref{#1}}
\newcommand{\partFig}[2]{Fig.~\hyperref[#1]{\ref*{#1}#2}}
\newcommand{\partFigure}[2]{Figure~\hyperref[#1]{\ref*{#1}#2}}

\newcommand{\latin}[1]{{\itshape #1}}
\newcommand{\eg}{\latin{e.\,g.}} 
\newcommand{\ie}{\latin{i.$\,$e.}} 
\newcommand{\vs}{\latin{vs}} 

\newcommand{\ue}{SUPA, School of Physics and Astronomy, University of Edinburgh, Peter Guthrie Tait Road, Edinburgh EH9 3FD, United Kingdom}
\newcommand{\hc}{The Hartree Centre, STFC Daresbury Laboratory, Warrington WA4 4AD, United Kingdom}

\begin{document}

\title{Diffusive Evaporation Dynamics in Polymer Solutions is Ubiquitous}

\author{Max Huisman}
 \email{m.huisman@sms.ed.ac.uk}
\author{Wilson C.\ K.\ Poon}
\affiliation{\ue}
\author{Patrick B.\ Warren}
    \affiliation{\ue}
    \affiliation{\hc}
\author{Simon Titmuss}
\author{Davide Marenduzzo}
 \email{davide.marenduzzo@ed.ac.uk}
    \affiliation{\ue}

\begin{abstract}
    Recent theory and experiments have shown how the buildup of a high-concentration polymer layer at a one-dimensional solvent-air interface can lead to an evaporation rate that scales with time as $t^{-1/2}$ and that is insensitive to the ambient humidity. Using phase field modelling we show that this scaling law constitutes a naturally emerging robust regime, Diffusion-Limited Evaporation (DLE). This regime dominates the dynamical state diagram of the system, which also contains regions of constant and arrested evaporation, confirming and extending understanding of recent experimental observations and theoretical predictions. We provide a theoretical argument to show that the scaling observed in the DLE regime occurs for a wide range of parameters, and our simulations predict that it can occur in two-dimensional geometries as well. Finally, we discuss possible extensions to more complex systems.
\end{abstract}


\maketitle

\section{Introduction}
Solvent evaporation from concentrated solutions or suspensions is an omnipresent phenomenon. The apparent simplicity of this process is deceptive: it is in fact ridden with complications due to the interaction between multiple components and the environment, leading to complex and fascinating dynamics. Such dynamics can in turn fundamentally alter the evaporative behaviour. Its understanding is therefore important for designing or controlling any process that involves drying. For instance, evaporative dynamics controls the application of paints and inks, where the formation of a defect-free skin upon drying is desired~\cite{DeGans}. It is also important in food preservation, where moisture content reduction increases shelf-life, but may also adversely affect flavour or texture~\cite{Erbay2010}. Due to such practical applications and fundamental interest in the dynamics of multi-component complex systems, the physics of evaporating polymer solutions and colloidal suspensions has inspired numerous investigations~\cite{Hennessy2017,DeGennes,Roger2016,Salmon2017,Rezaei2021}.

A number of these pertain to quasi-one-dimensional evaporation from the open end of a long capillary. In this geometry, the phase behaviour of aqueous lipid solutions can respond to varying ambient water activity $a_e$ (equivalently, relative humidity) in such a way as to render the evaporation rate practically independent of $a_e$~\cite{Roger2016}.  It is suggested that in this regime, the mass flux is diffusion-controlled, $J_{\text{evap}} \sim t^{-1/2}$. A similar regime was predicted theoretically by Salmon \etaltxt\ \cite{Salmon2017}. They argue that the evaporation-induced advective flux causes the growth of a concentrated `polarization layer' at the interface, leading to mass loss increasing with the square root of time, hereafter referred to as Diffusion-Limited Evaporation (DLE). However, the stability range of this DLE regime in parameter space was not explored because current calculations {\it assume} diffusional dynamics.

We have recently tested the predictions of Salmon \etaltxt, and observed a regime of $a_e$-independent evaporation in which the rate decreases as $t^{-1/2}$ \cite{Huisman2023}. We also found a transition from constant evaporation rate at early times to this $t^{-1/2}$ regime, consistent with the building up of a polymeric `polarisation layer'. The inclusion of elastic effects from the formation of a very thin `gelled skin' right at the air-solution interface~\cite{Ozawa2006,Okuzono2008} improves the agreement between theory and experiments.  

On general grounds, we expect that $J_{\text{evap}} \sim t^{-1/2}$ should only be one of at least three  dynamical regimes of mass loss in an evaporating polymer solution. At very low polymer concentration, we should approach pure solvent evaporation, where the mass loss $m(t)\sim t$, giving a constant evaporation rate. In the absence of any evaporative driving force, for instance when the air is saturated with solvent, we expect $m(t)\sim t^0$. So, we expect that DLE, where $J_{\text{evap}} \sim t^{-1/2}$ or $m(t)\sim t^{1/2}$, represents an intermediate regime. Surprisingly, however, this is the only behaviour observed experimentally to date at long times \cite{Roger2016,Roger2021,Huisman2023}. Why this is so is currently a puzzle.

Here we set up and investigate a one-dimensional continuum phase-field model for evaporation of a polymer solution, where the evaporation is hindered by the polymer. We indeed find three main evaporative regimes that transition into one another, with predictions for the diffusive regime that agree with other studies to date. Importantly, the DLE regime dominates the predicted state diagram of the system. We characterise the generic nature of this regime and propose an argument to explain why it is so pervasive. The physical picture that emerges is a simple one. A polymer layer grows from the interface into the bulk solution. When this layer becomes concentrated enough to act as a `porous plug', Darcy solvent flow through this layer is the rate limiting step, so that the evaporative dynamics becomes diffusive.
Finally, we show how the model can be extended to higher dimensions, or  to study more complex systems such as aerosol droplets, important in respiratory virus transmission, or multilayered paints and coatings. 

\section{Methods}
\subsection{A Phase Field Model for Evaporation}
Our phenomenological model consists of two phases, inside the drop and outside (= the atmosphere), connected through a continuous interface, with periodic boundary conditions (see SI for details). Two continuum fields are considered. 
The first of these is the order parameter $\phi$. In phase field modelling, the 
field $\phi$ is used to distinguish between two phases, for instance indicating a change in 
orientational order~\cite{Provatas2011}. In our model $\phi$ represents the total volume density of the droplet, that differentiates the droplet phase with a 
higher value of $\phi$, 
from the surrounding air, with a lower value of $\phi$. The second field, $c$, represents the concentration of polymer, and is in practice non-zero only in the droplet phase. Note that $\phi$ and $c$ have dimensionless units. Inside the droplet $\phi\simeq \phi_1$, and $c$ has some finite value, with $c=c_0<\phi_1$ initially. We follow the phase field convention that $\phi_1=1$ inside the drop, and note that therefore our model is fundamentally different from a two-fluid model, as $\phi+c \neq 1$. Outside the drop, there is no polymer ($c\simeq 0$), and the droplet phase field has some finite value $\phi\simeq \phi_0<\phi_1$: this phase represents ambient air, with $\phi_0$ the equivalent of the relative humidity.

\begin{figure*}[t!]
\includegraphics[width=\textwidth]{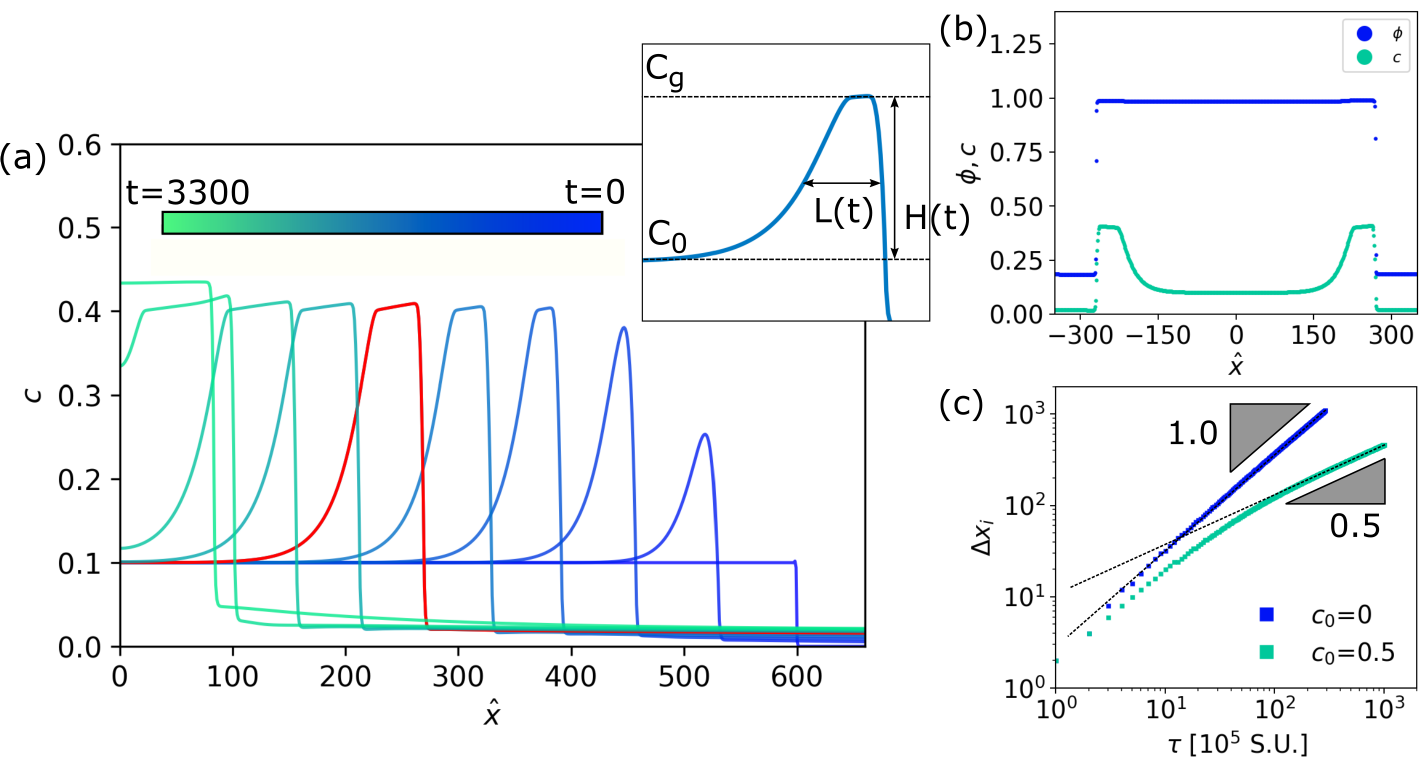}
\caption{\label{fig:1} Dynamic evolution of a unidirectional drying polymer-solvent system in 1D. \textbf{(a)} Time evolution of polymer concentration profiles $c$ in a 1D polymer-solvent slab during evaporation. The colorbar indicates the time corresponding to the concentration profiles. The $x$-coordinates are shifted from the simulation coordinates such that the middle of the droplet is located at $x=0$. Inset: schematic of the initial concentration $c_0$, gelation concentration $c_g$, width of the polymer layer $L(t)$ (taken as the Full Width Half Maximum) and the peak height of the polymer layer $H(t)$. \textbf{(b)} Snapshot of a typical simulation profile for $\phi$ and $c$, at time $t = 1800$, which corresponds to the \textcolor{red}{red} profile of $c$ in panel (a). The $x$-coordinates are shifted so that the middle of the droplet is located at $x=0$. \textbf{(c)} Evolution of the interface $\Dxi$ plotted over time, comparing a system with added polymer to a system of pure solvent. A log-log scaling is applied to highlight the different long time power-law behaviour. Simulation units (S.U.) for space $\Delta x_i$ and time $t$ can be converted to physical units by applying the scalings $L=2.5\times10^{-10}~$m and $T=6.3\times10^{-9}~$s, respectively, as presented in the main manuscript.} 
\end{figure*}

The dynamic evolution of these fields is governed by chemical potential gradients around the interfaces, stemming from a coupled free energy density $f(\phi,c)$. The two fields themselves are coupled through a convective term $\vg(\phi,c)$. Each phase field is described by a modified Cahn-Hilliard equation with an additional evaporative term~\cite{Lee2021}: 
\begin{subequations}
\label{eqs:field}
\begin{align}
    \label{eq:phase_field}
  &\frac{\partial \phi}{\partial t} + v_i \cdot \nabla \phi
  = \nabla \cdot \left[M_{\phi} \nabla \mu_{\phi} \right]\,,\\[3pt]
    \label{eq:c_field}
  &\frac{\partial c}{\partial t} + \nabla \cdot (v c)
  = \nabla \cdot \left[M_{c} (c) \nabla \mu_{c}\right]\,,
\end{align}
\end{subequations}
where note that $\phi$ is not conserved, whereas $c$ is conserved globally. This model requires that the total volume density inside the droplet, where $\phi=\phi_1$, and outside the droplet, where $\phi=\phi_0$, are essentially constant, such that $\phi$ is only lost around the interface where $\nabla \phi \neq 0$, \partFig{fig:1}{b}. A nearly constant value for $\phi$ in the bulk of different phases is common to phase field modelling \cite{Provatas2011}, and we suggest that this is also a reasonable assumption to describe the physics in our model, as dissolved polymers in solution have a comparable density to pure water, while the volume density of the water in the (well-mixed) ambient air is approximately constant and should only vary in a small region right next to the interface of the evaporating solution. The constant volume density inside the droplet $\phi_1$ also means that the internal droplet dynamics are conserved. Furthermore, in \Eqs{eqs:field} $M_\phi$ is the mobility of the phase field $\phi$, which we take as a constant, while the polymer mobility is concentration dependent, $M_{\text{c}}(c) = \frac{M_0}{1+\beta c}$. We note that using a constant polymer mobility leads to qualitatively similar evaporation dynamics, which was also observed in \cite{Salmon2017}. 

The quantity $v_i$ is the interfacial velocity of the evaporating droplet. Physically, the driving force for evaporation is the interfacial water activity difference \cite{Cussler1997}. In the absence of polymer, it was suggested that this can be represented by a gradient in $\phi$ \cite{Lee2021}, leading to $v_i \sim \nabla \phi$. In our model, as the water activity inside the droplet reduces with increasing presence of polymer, an increasing polymer concentration $c$ should reduce evaporation. Therefore we take the phenomenological expression $v_i = \gamma \nabla (\phi - (\gamma'/\gamma) c)$, with the parameters $\gamma$ determining the relative importance of evaporation to the phase fields and $\gamma^\prime$ the contribution of $c$ to reducing evaporation. In the droplet with binary composition, $\nabla (\phi - (\gamma'/\gamma) c)$ is the effective solvent gradient leading to evaporation. We note that this expression is only valid for evaporation if $\phi \geq (\gamma'/\gamma) c$, and that the expression is similar to those used in other models in terms of water activity \cite{Salmon2017} or partial pressure \cite{OKAZAKI1974}. Using an alternative expression, where the interface velocity also depends on the local phase field as $v_i' = \phi \times \gamma \nabla (\phi - (\gamma'/\gamma) c)$, gives similar results in the same parameter range, with a slight renormalization, as we show in the Supplemental Information. Finally, $v=-v_i$ in Eq.~(\ref{eq:c_field}) is the water velocity which advects the polymer towards the interface. Inside the droplet $\nabla c \neq 0$, so in Eq.~(\ref{eq:c_field}) there exists a convective flux of polymer towards the interface, which is the evaporating solvent flux to the interface, compressing the polymer \cite{Meng2016}. This is not the case in Eq.~(\ref{eq:phase_field}), as inside the droplet $\nabla \phi = 0$, hence the dynamics there is purely diffusive.

The term driving droplet evaporation is therefore $v_i \cdot \nabla \phi$ at the droplet-air interface, which leads to droplet shrinking. Interestingly, in the absence of $c$, this has the form of a square gradient term, similar to the key nonlinearity in the Kardar-Parisi-Zhang (KPZ)~\cite{Kardar1986} equation, but with the opposite sign with respect to the one normally considered for growing interfaces.

The local chemical potential $\mu$ is derived from the free energy density $f$ as $\mu_{\phi} = \frac{\delta F}{\delta \phi} =  \frac{\partial f}{\partial \phi} - \nabla \cdot \frac{\partial f}{\partial \nabla \phi}$ for the solvent and $\mu_{\text{c}} = \frac{\delta F}{\delta c}  = \frac{\partial f}{\partial c} - \nabla \cdot \frac{\partial f}{\partial \nabla c}$ for the polymer. We note that in phase field modelling, it is common practice to use a free energy landscape $f(\phi,c)$ of the phase fields $\phi$ and $c$ to generate physically appropriate chemical potential terms that describe the evolution of the phase fields in \Eq{eq:phase_field} and \Eq{eq:c_field} \cite{Provatas2011}. This definition of $\phi$ and $c$ as phase fields is different compared to theoretical treatments such as Landau-Ginzburg theory, that use thermodynamic density fields. We use a Landau-like free energy density for the phase fields $\phi$ and $c$ as 
\begin{equation}
\label{eq:free_energy_density}
\begin{split}
  f(\phi,c) &= \frac{a_1}{4}(\phi - \phi_0)^2(\phi - \phi_1)^2
  + \frac{\kappa_\phi}{2}|\nabla \phi|^2 \\[3pt]
    &{} + \frac{\kappa_c}{2}|\nabla c|^2  - \frac{a_0}{2}\phi^2 c^2 + g(x)\frac{a_{\text{2}}}{2}c^2 \\[3pt]
    &{} +  G(x)\frac{K_{\text{g}}}{2}(c-c_{\text{g}})^{2} + \frac{b_0}{2}c^2 + \frac{b_1}{4}c^4 \,.
\end{split}
\end{equation}
The first term ensures that the system separates into a solvent-rich droplet $\phi_1$ with polymer ($c > 0$) and a surrounding vapor phase $\phi_0$ that contains no polymer ($c = 0$), so that $\phi_0$ represents the solvent concentration. $\kappa_\phi$ and $\kappa_c$ determines the bare surface tension of the droplet and polymer. The phenomenological term $-\frac{a_0}{2}\phi^2 c^2$ is necessary to confine the polymer into the droplet interior, and its form is chosen in analogy to other types of phase field modelling~\cite{Mueller2019}; similar to those types of models, we expect that different forms favouring polymer confinement should lead to qualitatively similar results. The term $g(x)\frac{a_{\text{2}}}{2}c^2$ penalises the transfer of polymer across the interface, where $g(x)=\Theta(\phi(x)-\frac{\phi_1 - \phi_0}{2})$ is an indicator function defined in terms of the Heaviside $\Theta$ such that $g=0$ if $\frac{\phi_1 - \phi_0}{2}>\phi(x)$ and $g=1$ otherwise. When $c$ is high enough to induce gelation, a permanent elastic stress develops, increasing the osmotic pressure and thereby the chemical potential \cite{Leibler1993}. We approximate the bulk osmotic modulus contribution with $G(x)\frac{K_{\text{g}}}{2}(c-c_{\text{g}})^2$ \cite{Okuzono2008}, where $K_{\text{g}}$ is a (constant) bulk osmotic modulus, $c_{\text{g}}$ is the gelation concentration of the polymer, and $G(x)=\Theta(c(x)-c_{\rm{g}})$ is another indicator function defined again in terms of the Heaviside $\Theta$. The remaining terms represent the virial coefficient for polymer diffusion ($b_0$) and excluded volume effects of the polymer~($b_1$). We note that \Eq{eq:free_energy_density} should be accordingly adjusted to be used for modelling other systems than evaporating polymer-water droplets.

From \Eq{eq:free_energy_density}, the system spontaneously phase-separates into a droplet phase and an environment phase, independent of initial conditions. We tested this by initializing the system with a sinusoidal variation of $\phi$ and $c$ with $x$, see Supplemental Information, and found that this system quickly phase-separates and stabilizes into the general profiles seen in \partFig{fig:1}{b}.

Finally, we note that another way to approach the evaporation problem could be to have a two-field (or Landau-Ginzburg) model, with one field for the interior of the droplet and another for the exterior, which are joined by a moving boundary condition. However, the implementation of such a model is non-trivial. In our implementation of the single phase-field model the moving boundary, i.e. the interface, is an emergent feature from the underlying physics. A numerical drawback of our implementation is that a fine-grained grid is required to resolve the interface structure. For one-dimensional and radially symmetric problems this is not an issue, but such limitations might become more relevant for systems with increased spatial heterogeneity. Another numerical cost is the need to construct a sufficiently elaborate phase-field model that captures all the relevant physics, which leads to a relatively large amount of numerical parameters in the governing equations. This is, however, expected to be similar compared to moving boundary problems, where multiple numerical parameters are also likely to be required to correctly describe the physics in the system. For our phase-field model, we have thoroughly explored the relevant parameter spaces, for which our findings are robust.

\section{Results and Discussion}
\subsection{Unidirectional Drying in 1D.} 
We solve a 1D version of our model with periodic boundary conditions using the computational procedure provided in the Supplemental Information, and defining the origin of the $x$ coordinate to be in the middle of the droplet. For the case  of  $\phi_1=1$, $c_0=0.1$, $\phi_0=0.2$ and $\gamma'/\gamma=1.50$, \partFig{fig:1}{a} shows the polymer $c(x,t)$ in the right half ($x > 0$) of the droplet at a series of time points, while \partFig{fig:1}{b} plots the profiles of $\phi$ and $c$ in the full droplet at time $t = 1800$. These results agree qualitatively with previous work \cite{Ozawa2006}. As the drop shrinks, a peak in $c(t)$ develops just within the interface -- a `polarisation layer'. When the peak height, $H(t)$, reaches $c_{\text{g}}$, the peak stops growing and flattens into a plateau of increasing width, $L(t)$ (\partFig{fig:1}{a} inset), until the concentration in the droplet is homogeneous and the droplet continues to shrink slowly. We note that outside the droplet the polymer concentration is not exactly $c=0$, indicating minor leakage of polymer into the environment. Such minor leakage does not significantly affect the overall evaporation dynamics.

To quantify evaporative dynamics, consider the position of the interface $x_{\rm i}$, taken to be the position of the peak in $c(t)$. \partFig{fig:1}{c} shows a log-log plot of $\Delta x_{\rm i} (t) = x_{\rm i} (t) - x_{\rm i}(0) = m(t)$ by mass conservation if only solvent leaves the interface. For pure solvent ($c_0=0$), $\Delta x_{\rm i} \sim t$, so that $J \sim \ddt{(\Delta x_{\rm i})}$ is constant, which is a well-known result~\cite{Cussler1997}. A solvent-polymer mixture ($c_0=0.5$, with $\phi_0=0.35$ and $\gamma'/\gamma=1.50$) behaves differently. After an initial linear regime, the evaporation slows down and approaches a steady state where $\Delta x_{\rm i} \sim t^{1/2}$ and $J \sim t^{-1/2}$, as found by experiments~\cite{Roger2016,Huisman2023} and theory~\cite{Salmon2017}. 

\begin{figure}[t!]
\includegraphics[width=0.48\textwidth]{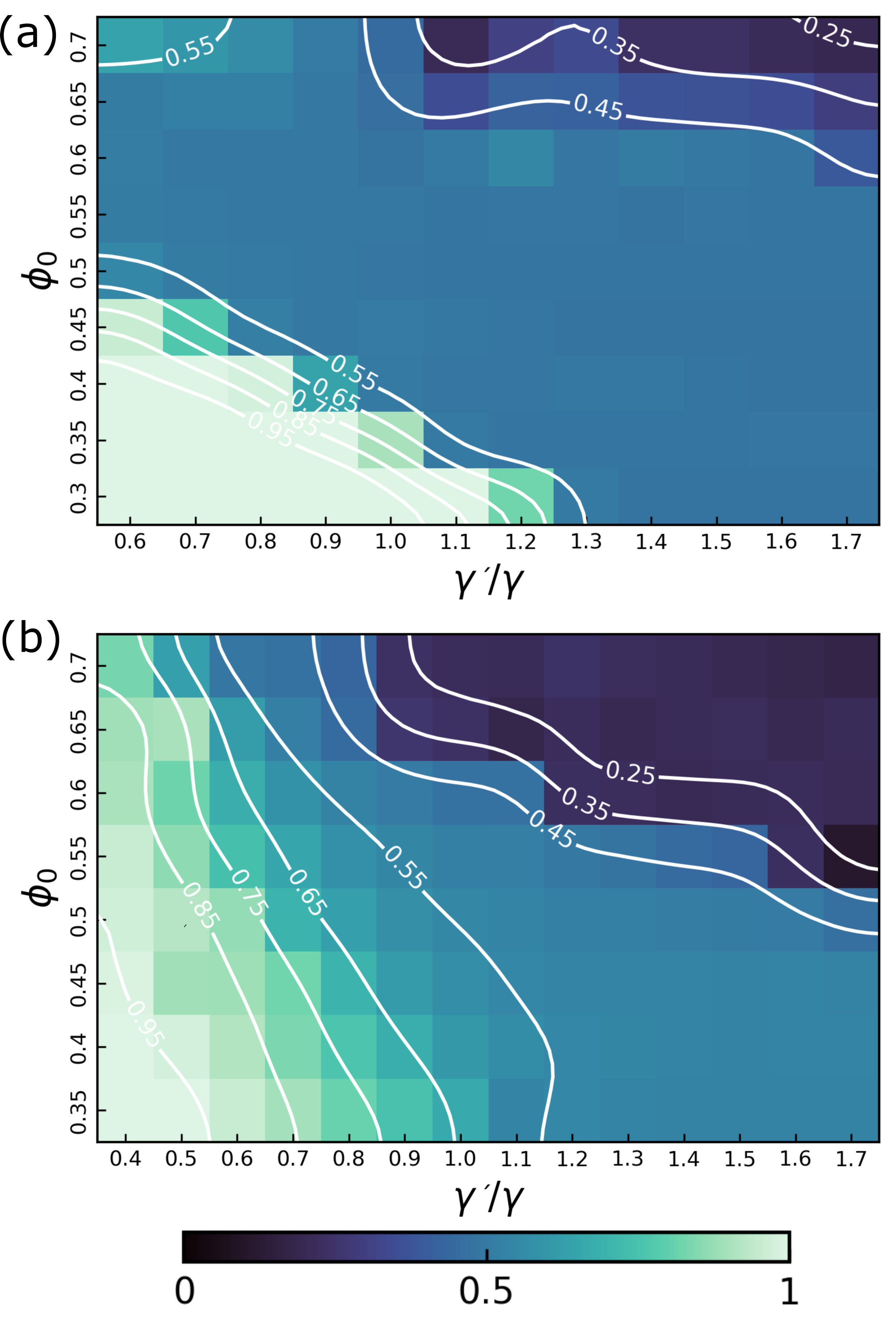}
\caption{\label{fig:2} State diagrams of the $\alpha$ exponent in $m(t) \sim t^{\alpha}$ with varying $\phi_0$ and $\gamma'/\gamma$, in a system with $c_0=0.5$, using \textbf{(a)} a Landau expansion for the polymer free energy $f_{\text{Landau}}(c) = \frac{b_0}{2}c^2 + \frac{b_1}{4}c^4$ and \textbf{(b)} a Flory-Huggins free energy $f_{\text{FH}}(c) = b_0 (1-c)\ln{(1-c)} + \chi c(1-c)$. Countour lines (white) are added to highlight the transition between the different evaporation regimes.} 
\end{figure}

\subsection{Stability of the DLE Regime.} To assess the relative stability of the $\Delta x_{\rm i} \sim t$ and $\sim t^{1/2}$ regimes and explore the possibility of other forms of scaling, we scan two parameters. 
The first, $\gamma'/\gamma$, regulates the extent to which $c$ reduces the convective evaporation speed -- recall $v_i = \gamma \nabla (\phi - (\gamma'/\gamma) c)$. The second is the phase field value outside the droplet, $\phi_0$, which is equivalent to relative humidity and governs the evaporative driving force. 

For the exponent $\alpha$ in $m(t)\sim t^\alpha$, the state diagram in \partFig{fig:2}{a} displays three dynamical regimes separated by relatively sharp boundaries, as indicated by the white contour lines. We identify the pure solvent-like regime ($\alpha \to 1$) in the bottom left corner of the state diagram and the arrested evaporation regime ($\alpha \to 0$) in the top right corner of the state diagram. The DLE regime where $\alpha \approx 0.5$ occupies the largest region in the state diagram, and is therefore the most robust. This is consistent with the fact that diffusive dynamics is the behaviour typically reported in experiments to date.

\begin{figure*}[t]
\centering
\includegraphics[width=\textwidth]{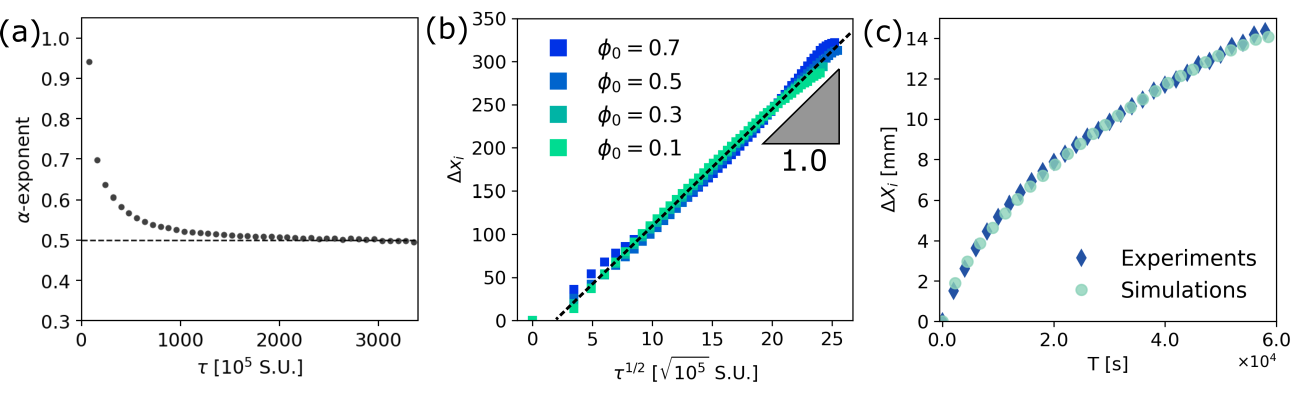}
\caption{\label{fig:3} Diffusion-Limited Evaporation (DLE) after long times that is independent of the external driving force. \textbf{(a)} Evolution of the exponent $\alpha$ in $m(t) \sim t^{\alpha}$ over time steps $\tau$ in a system with $\phi_0=0.35$ and $\gamma'/\gamma=1.5$, settling on a time exponent $\alpha = 0.5$. $\alpha$ is found through power law fitting of $m(t)=B t^{\alpha}$ using $\alpha = {\leibnizd{\,\ln m(t)}}/{\leibnizd{\,\ln t}}$. \textbf{(b)} Evolution of $\Dxi$ over the square root of time $\tau^{1/2}$, showing independence of the external driving force $\phi_0$. Systems shown are $\phi_0=0.7$ with $\gamma'/\gamma=0.415$, $\phi_0=0.5$ with $\gamma'/\gamma=1.00$, $\phi_0=0.3$ with $\gamma'/\gamma=1.85$ and $\phi_0=0.1$ with $\gamma'/\gamma=2.75$. In (a) and (b) simulation units (S.U.) for space and time are $L=2.5\times10^{-10}~$m and $T=6.3\times10^{-9}~$s, respectively. \textbf{(c)}. Evolution of the interface in dimensional units, scaled to experimental data as $\Delta X_i = [X] \Delta x_i$, with $[X] = 4.5 \times 10^{-5}~$m, and $T = [T] t$, with $[T]=9 \times 10^{-4}~$s. Simulation dataset is from a system with $\phi_0=0.5$ with $\gamma'/\gamma=1.00$. Experimental dataset at 50\% relative humidity is reproduced from \cite{Huisman2023}.} 
\end{figure*}

To understand the stability of the $\Delta x_{\rm i} \sim t^{1/2}$ regime, note that \Eqs{eqs:field} become diffusion equations in the limit $v_i \to 0$~\cite{Salmon2017}. The system cannot start in this regime, but can only approach it asymptotically: having
$\gamma'/\gamma \sim \nabla_{\rm i} \phi/ \nabla_{\rm i}  c$ to give $v_i \to 0$ (where $\nabla_{\rm i} \equiv$ gradient at the interface) means no evaporation in the first place. We therefore need $\gamma^\prime/ \gamma \lesssim \nabla_{\rm i} \phi/ \nabla_{\rm i} c$ to confer a finite initial evaporation rate, which then decreases with time as interfacial polymer accumulates and the system approaches the diffusive regime ($\alpha = 0.5$) asymptotically, \partFig{fig:3}{a}. How fast this happens depends on the effectiveness of interfacial polymer in reducing evaporation, which is controlled by $\gamma^\prime$. 

As $\gamma'/\gamma$ drops, this effectiveness decreases, requiring a larger polarisation layer that takes longer to establish to approach the diffusive regime. So, for finite system size and observation time, there exists a  $(\gamma'/\gamma)_{\rm min} = \epsilon$ below which the system will not cross over to $\Delta x_{\rm i} \sim t^{1/2}$ behaviour. We expect $\epsilon$ to increase with the polymer mobility, $M_{\rm c}$: more mobile polymers require a longer time to build up a large enough polarisation layer to slow evaporation. On the other hand, a stronger driving force for evaporation due to lower external solvent concentration, $\phi_0$, requires the polymer to be more effective in reducing evaporation for diffusive behaviour to emerge; so, $\epsilon$ should decrease with increasing $\phi_0$, as observed, \Fig{fig:2}.

We establish that the scaling results from our modelling approach are insensitive to the exact implementation of the polymer free energy, by replacing the polymeric contribution from a Landau expansion $f_{\text{Landau}}(c) = \frac{b_0}{2}c^2 + \frac{b_1}{4}c^4$ to the free energy density (\Eq{eq:free_energy_density}) with the Flory-Huggins mean field expression $f_{\text{FH}}(c) = b_0 (1-c)\ln{(1-c)} + \chi c(1-c)$, assuming a large degree of polymerization $N \gg 1$, see Supplemental Information for more details. The resulting state diagram for varying $\gamma^\prime/\gamma$ and $\phi_0$ in \partFig{fig:2}{b} is dominated by a large region where $m(t) \sim t^{0.5}$, indicating DLE, and the transition of long-times exponents from $\alpha=1$ to $\alpha = 0.5$ and $\alpha = 0$ in $m(t) \sim t^\alpha$ is qualitatively similar between \partFig{fig:2}{a} and \partFig{fig:2}{b} upon increasing $\gamma`/\gamma$ and $\phi_0$. We note that the boundaries between the evaporation regimes in \partFig{fig:2}{b} are less well-defined, which is shown by comparing a line in the state diagrams in the Supplemental Information. We attribute this effect to the divergence of the term $\sim \ln{(1-c)}$ in $f_{\text{FH}}(c)$ for $c \to 1$, which leads to increased spreading and therewith slight variations in the settling interface concentration $c_i<1$, that are likely to depend on the evaporation driving force, set by $\phi_0$.

To understand the physics underpinning the late-stage, diffusive regime, note that by this stage, the chemical potential of the solvent (at partial pressure $p$) just inside the interface (where the polarisation layer is at its most concentrated), $\mu(p)|_{{x}_{\text{i}}}$, has nearly equilibrated with that of the solvent vapour outside, and so is nearly constant. At the same time, the solvent chemical potential in the middle of the droplet, $\mu(p)|_{{x}=0}$, is also constant. So, there is a constant osmotic pressure difference driving solvent flow through the polarisation layer towards the interface, $\Delta p = p|_{{x}_{\text{i}}}-p_{{x}=0} < 0$ (because $\mu(p)|_{x_{\rm i}} < \mu(p)|_{{x}=0}$). Treating the growing polarisation of thickness $L(t)$ as a porous medium of Darcy permeability $k$ implies the solvent flux $J = -\frac{k}{\eta L(t)}\Delta p$, with $\eta$ the solvent viscosity. In the DLE regime in steady state, this flux is exactly balanced by the evaporative flux, which mass conservation requires to be $\ddt{L}$, so that we have $\ddt{L} \sim L^{-1}$, or $L(t)\sim t^{1/2}$, as also found in a recent theory~\cite{Talini2023} as well in our numerics (see SI II.A).  Darcy's law then implies $J \sim t^{-1/2}$, consistent with our finding that $\Delta x_{\rm i} \sim t^{1/2}$ at long times (recall that $J \sim \ddt{(\Delta x_{\rm i})}$), \partFig{fig:1}{c}.

For this physical argument to hold, we require that the rate of advective polymer accumulation, giving rise to the (growing) polarisation layer, dominates polymer diffusion which counteracts this buildup, 
\ie~the P\'eclet number $\Pe\equiv\frac{v_i L(t)}{M}\gg1$. Equivalently, this means that the diffusion is slow compared to the evaporation rate. While this is true at short times, it seems at least at first sight that this condition will be broken as late times where na\"\i{}vely one might expect $\Pe\to0$ as $v_i\to 0$. However, in the DLE regime, our argument suggests $v_i \sim \dot{m}(t) \sim t^{-1/2}$, $L(t) \sim t^{1/2}$. So in fact at late times $\Pe\sim t^0$, provided that $M_{\text{c}} = \frac{M_0}{1+\beta c} \simeq\const$. The latter is a fair approximation for systems with $\beta \sim 0.1$--1.0 and $c \sim 0.1$--1.0. Our physical argument for the diffusive regime at long times is therefore self consistent, because the balance between advection and diffusion remains constant even as $v_i \to 0$. For a final check, we find $\Pe\approx 10^{2} \gg 1$ for a typical system in the DLE regime ($\gamma'/\gamma=1.50$, $\phi_0=0.30$), so that, indeed, advection near the interface dominates over diffusion. 

\begin{figure*}[t]
\includegraphics[width=\textwidth]{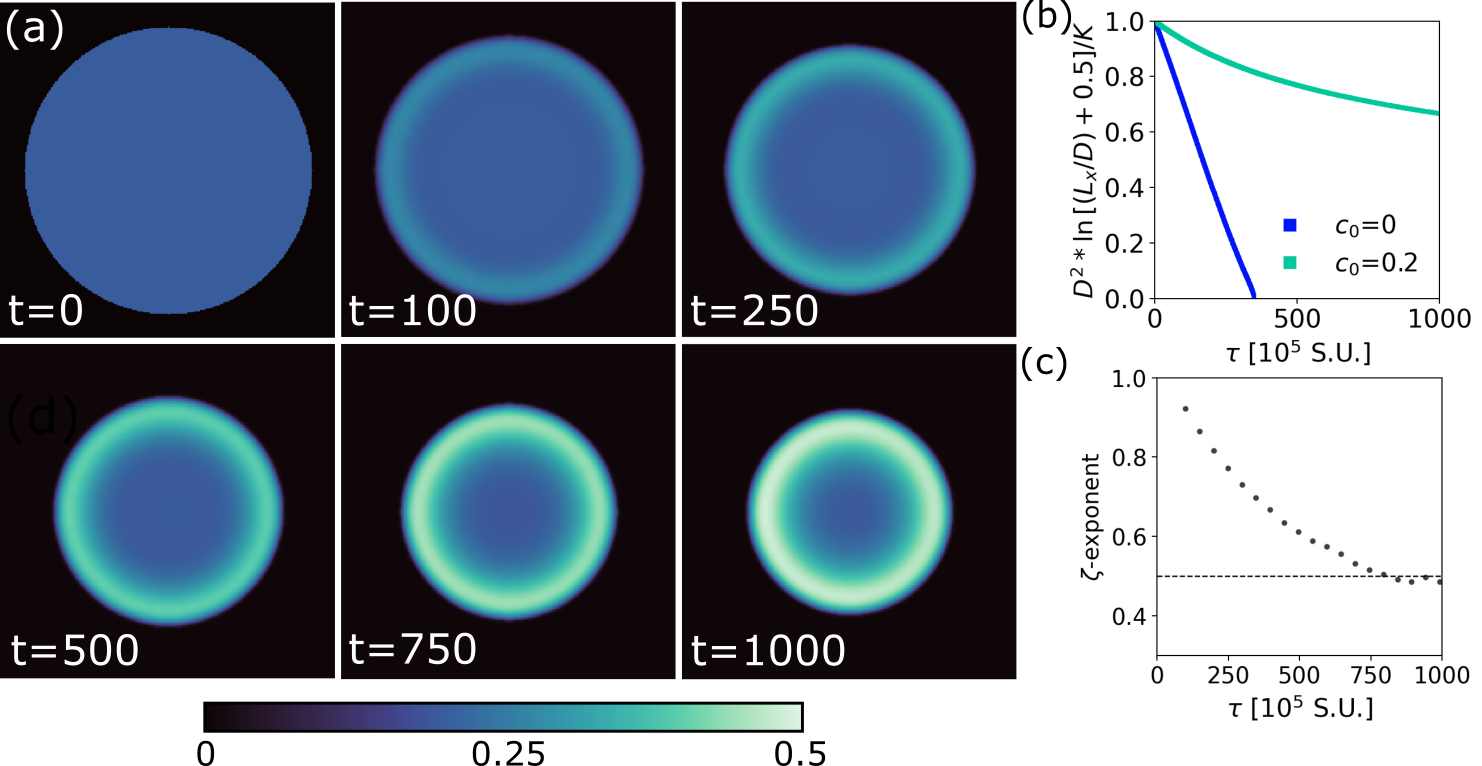}
\caption{\label{fig:4} Dynamic evolution of evaporating droplets in 2D. \textbf{(a)} Snapshots of the time evolution of polymer concentration $c$ in a 2D evaporating droplet, with $c_0=0.2$, $\phi_0=0.35$ and $\gamma'/\gamma=3.0$. \textbf{(b)} Evolution of the log-corrected area $D^{2}\ln{[(L_{\text{x}}/D)+0.5]}/G$ of the system in (a) plotted over time, comparing a system with added polymer to a system of pure solvent. \textbf{(c)}  Evolution of the exponent $\zeta$ in $D^{2}\ln{[(L_{\text{x}}/D)+0.5]} \sim t^{\zeta}$ over  timesteps $\tau$ in a system with $\phi_0=0.35$ and $\gamma'/\gamma=3.0$, settling on a time exponent $\zeta = 0.5$. $\zeta$ is found through power law fitting using $\zeta = {\leibnizd\ln{\left[1 -(D^{2})\ln{[(L_{\text{x}}/D)+0.5]}/G \right]}}/{\leibnizd\ln{t}}$. In (b) and (c) simulation units (S.U.) for space and time are $L=2.5\times10^{-10}~$m and $T=6.3\times10^{-9}~$s, respectively.} 
\end{figure*}

Considering the concentration dependence of our results, we identify the limiting evaporation regimes of $m(t) \sim t^{\alpha}$ for a varying initial concentration $c_0$. For a system where $c_0 \to 0$, we expect that DLE never occurs and the system evaporates at a constant rate ($\alpha \to 1$). On the other hand, if $c_0$ is sufficiently high that the system is already at thermodynamic equilibrium with the environment, evaporation never occurs ($\alpha \to 0$). For any other value of $c_0$, we expect to observe the behaviour where $0<\alpha<1$, which approaches $\alpha\to0.5$ as long as $\Pe\gg 1$ and the system is sufficiently large that a polarization layer can form at the solution-air interface.

In the diffusive regime, it was previously predicted~\cite{Salmon2017} that the mass loss rate should be independent of the external driving force. For water evaporation into air, the driving force is the relative humidity $a_e$ \cite{Cussler1997}, which for us is `tuned' by $\phi_0$. Plotting $m$ \vs\ $t^{1/2}$ whilst varying $\phi_0$ and $\gamma'/\gamma$, \partFig{fig:3}{b}, shows that this is indeed the case in our model.

This is a direct consequence of the fact that interfacial polymer concentration has reached a constant value, $c_{\rm{g}}$,  so that it is the polymer concentration gradient in the polarisation layer rather than the external humidity that drives water transport. In the theory of Salmon \etaltxt~\cite{Salmon2017}, the same physics emerges due to the sharp fall in water activity at high polymer concentrations, so that the late stage interfacial polymer concentration varies very little over a broad range of external water activities. In both cases, the humidity independence is necessarily correlated with the emergence of DLE. However, such correlation is not logically necessary. In our earlier experiments, the formation of a thin polymer skin at the solution-air interface due to rapid adsorption also gives rise to humidity-independent evaporation, but without a porous polarisation layer, the dynamics is not diffusive~\cite{Huisman2023}.

\subsection{Simulation Units and Comparison to Experimental Systems.} We find the simulation units in time ($T$) and space ($L$) of our system, by comparing the simulation parameters in \Eq{eq:free_energy_density} to measured physical quantities of a water-PVA solution, for which measurements have been performed in experiments where a solution evaporates unidirectionally \cite{Huisman2023}. Using the units of the simulation parameters from Table S1 in the Supplemental Information, we obtain a set of equations for (1) the surface tension $[\gamma] = \sqrt{8\kappa_\phi a_1/9} \times E/L^2$ \cite{Cates2018}, (2) the diffusion coefficient $[D] = M_0 b_0 \times L^2/T$ and (3) the osmotic pressure $[\pi] = b_0 \times E/L^3$, where $E$ is the unit of energy. Using $[\gamma] \approx 0.07$, $[D] \approx 10^{-11}$ \cite{OKAZAKI1974} and $[\pi] \approx 10^7~$Pa \cite{Bacchin2021}, we find $L\approx2.5\times10^{-10}~$m, $T\approx6.3\times10^{-9}~$s and energy unit $E \approx 1.56 \times 10^{-20}~$J. Finally, we calculate a dimensional evaporation rate from the slope of a simulation without polymer ($\phi_0=0.5$, $c_0=0$) as $V_{ev} \approx 2 \times 10^{-5} L/T \approx 0.8~\mu$m/s, which is comparable to measurements in \cite{Huisman2023}.

To establish that our simulations are representative of experimental data, we map our results to the data from \cite{Huisman2023} in \partFig{fig:3}{c}. We find excellent agreement between these two systems by scaling our simulation data to dimensional units as $\Delta X_i = [X] \Delta x_i$ in meters [m], where $[X] = 4.5 \times 10^{-5}~$m, and $T = [T] t$ in seconds [s], where $[T]=9 \times 10^{-4}~$s. It should be noted that these scalings are much higher than the simulation units in the system, which can be attributed to the experimental system being much larger than we can reasonably simulate, but that the ratios $L/T \approx 0.05$ and $[X]/[T] \approx 0.043$ are consistent.

\if{
We assess the validity of this approach by comparing the simulation mobility to an experimental mobility. In simulations, the polymer mobility scales approximately as $[M] \sim 0.1 \frac{[X]^2}{[F][T]}$, where the factor 0.1 comes in because we used $M_0 \approx 0.1$ and $[F]$ is a scale for the free energy density. For a neutral polymer in a good solvent $[F]$ should scale approximately as $\kT$ per correlation volume $\xi^3 (\sim R_H^3)$. Using $R_H \sim 10~$nm, we find $[F] \sim 10^3 \frac{\kT}{R_H^3} \approx 4 \times 10^{6}~$Pa, where a factor $10^3 = \frac{1}{10^{-3}}$ was added because $b_0 \approx 10^{-3}$ in \Eq{eq:free_energy_density}, and $[M] \sim 5 \times 10^{-14}~$m$^2$/(Pa s). Comparing this to the experimental mobility $M_{\text{exp}} \sim \frac{D_{\text{SE}}}{\kT / R_H^3} \approx 2\times 10^{-15}~$m$^2$/(Pa s), with Stokes-Einstein diffusion coefficient $D_{\text{SE}} \sim 10^{-11}$ m$^2$/s and $R_H \sim 10~$nm, we find that the mobility values are only about one order of magnitude apart, confirming the applicability of this approach and our framework.
}\fi

\subsection{Extensions of the Model to more Complex Systems.} One of the major advantages of our phase field model is the ease with which it can be adapted to more complex systems and used to explore higher dimensions. \partFig{fig:4}{a} shows the evolution of a 2D evaporating drop geometry for a system with $c_0=0.2$, $\phi_0=0.35$ and $\gamma'/\gamma=3.0$. As in 1D, a concentrated polymer layer forms at the droplet-air interface over time. 

For 2D or 3D droplets of diameter $D$ evaporating in an unconfined environment, $D^{2} \sim t$, so that the evaporation rate is constant~\cite{Fei2022}. However, solving the same problem in a confined system with a constant solvent chemical potential imposed at the system's boundary leads to deviations from this `$D^2$ law'. If viscous and buoyancy effects can be neglected, theory~\cite{Fei2022} predicts that in a 2D finite system of this kind,
\begin{equation}
    \label{eq:D_law}
    D^{2}\ln{[(L_{\text{x}}/D)+0.5]}/G = 1 - Ct\,,
\end{equation}
where $L_{x}$ is the size of one axis of the system, $C$ is a constant and $G=D_0^{2}\ln{[(L_{\text{x}}/D_0)+0.5]}$. Over the course of the simulations that give the results shown in \Fig{fig:4}, the value of $\phi$ at the (periodic) boundaries increases by $\lesssim 2\%$, so that we may expect \Eq{eq:D_law} to hold to a good approximation.

Our data for the evaporation of a pure solvent droplet, \partFig{fig:4}{b}, (blue; $c_0 = 0$) indeed agree with \Eq{eq:D_law}, as apparent from the approximately linear evolution of $D^{2}\ln{[(L_{\text{x}}/D)+0.5]}/G$ with $t$. However, in a system with added polymer, deviations are observed, \partFig{fig:4}{b} (green, $c_0=0.2$), which recall \partFig{fig:1}{c}. We therefore fit these data to a power law with a running exponent $\zeta$, $D^{2}\ln{[(L_{\text{x}}/D)+0.5]} \sim t^{\zeta}$. 
The resulting $\zeta$, \partFig{fig:4}{c}, clearly recalls \partFig{fig:3}{a} for the 1D case. 

So, while a full study of the 2D case is beyond our scope, it seems reasonable to surmise that the state diagram in this case should also display a robust diffusive regime, provided that $\Pe$ is high enough. Previous experiments have demonstrated ambient humidity independent evaporation of droplets containing large glycoproteins~\cite{Vejerano2018}, which is consistent with our surmise. 

\section{Conclusions and Outlook}

In summary, we have applied a phase field modelling approach to study the evaporative dynamics of a polymer-solvent mixture. Whilst our approach is phenomenological, rather than being derived from rigorously coarse-graining a microscopic theory, we expect it to describe the system in a qualitatively accurate way, in line with previous work on phase fields. Our key result is that the DLE regime, where evaporation rate decays with time as $t^{-1/2}$, is a robust dynamical regime found over a range of parameter values. We rationalise this scaling with a simple mathematical and physical argument, according to which the $-1/2$ exponent is due to a diffusive growth of the polymer layer, and a Darcy flow of the solvent due to the ensuing pressure difference close to the droplet-air interface. For this argument to be self-consistent, the P\'eclet number $\Pe$ should remain high and nearly constant, to achieve a non-equilibrium steady state where advection dominates over diffusion at all times. We show that this requirement is, perhaps surprisingly, indeed met. 

Our model is in quantitative agreement with previous theoretical and experimental results, including a near-independence of evaporation rates on relative humidity in the DLE regime. Such agreement gives confidence for applying this phase field model to study solvent and solute transfer in more complex systems and geometries. We have shown preliminary results for the evaporation of a 2D droplet to demonstrate this potential. Possible future applications include dissolution processes (\eg, making instant coffee) and the drying of multilayers involving multiple solvents and solutes (\eg, oil paintings~\cite{Duivenvoorden2023}). In the latter case, our approach has the added advantage that no assumptions need to be made on the phase of layers and/or the location of the interface over time. 

\section*{Acknowledgements}

Funding was provided by the University of Edinburgh. For the purpose of open access, the author has applied a Creative Commons Attribution (CC BY) licence to any Author Accepted Manuscript version arising from this submission.

\section*{References}

\providecommand*{\mcitethebibliography}{\thebibliography}
\csname @ifundefined\endcsname{endmcitethebibliography}
{\let\endmcitethebibliography\endthebibliography}{}

\end{document}


\newcommand{\kB}{k_{\text{B}}}
\newcommand{\kT}{\kB T}

\newcommand{\Dxi}{\Delta x_{\text{i}}}

\newcommand{\vg}{v}

\newcommand{\leibnizd}{\mathrm{d}}
\newcommand{\dt}{\leibnizd{t}}
\newcommand{\ddt}[1]{\leibnizd{#1}/\dt}

\newcommand{\Pe}{\mathrm{Pe}}
\newcommand{\const}{\mathrm{constant}}

\newcommand{\Eq}[1]{Eq.~\ref{#1}}
\newcommand{\Eqs}[1]{Eqs.~\ref{#1}}
\newcommand{\Fig}[1]{Fig.~\ref{#1}}
\newcommand{\partFig}[2]{Fig.~\hyperref[#1]{\ref*{#1}#2}}
\newcommand{\partFigure}[2]{Figure~\hyperref[#1]{\ref*{#1}#2}}

\newcommand{\naive}{{na\"\i{}ve}}

\newcommand{\Peclet}{P\'{e}clet}

\newcommand{\latin}[1]{{\itshape #1}}
\newcommand{\cf}{\latin{cf.}} 
\newcommand{\etc}{\latin{et\,c}} 
\newcommand{\eg}{\latin{e.\,g.}} 
\newcommand{\ie}{\latin{i.$\,$e.}} 
\newcommand{\vs}{\latin{vs}} 
\newcommand{\etaltxt}{\latin{et al.}} 

\newcommand{\rev}[1]{\textcolor{blue}{#1}}

\title{Diffusive evaporation dynamics in polymer solutions is ubiquitous - Supplemental Information}

\author{Max Huisman}
\affiliation{SUPA, School of Physics and Astronomy, University of Edinburgh, Peter Guthrie Tait Road, Edinburgh EH9 3FD, United Kingdom}
\author{Wilson C. K. Poon} 
\affiliation{SUPA, School of Physics and Astronomy, University of Edinburgh, Peter Guthrie Tait Road, Edinburgh EH9 3FD, United Kingdom}
\author{Patrick B. Warren}
\affiliation{SUPA, School of Physics and Astronomy, University of Edinburgh, Peter Guthrie Tait Road, Edinburgh EH9 3FD, United Kingdom}
\affiliation{The Hartree Centre, STFC Daresbury Laboratory, Warrington WA4 4AD, United Kingdom}
\author{Simon Titmuss}
\affiliation{SUPA, School of Physics and Astronomy, University of Edinburgh, Peter Guthrie Tait Road, Edinburgh EH9 3FD, United Kingdom}
\author{Davide Marenduzzo}
\affiliation{SUPA, School of Physics and Astronomy, University of Edinburgh, Peter Guthrie Tait Road, Edinburgh EH9 3FD, United Kingdom}

\maketitle

\begin{figure}[h]
\centering
\includegraphics[width=0.5\textwidth]{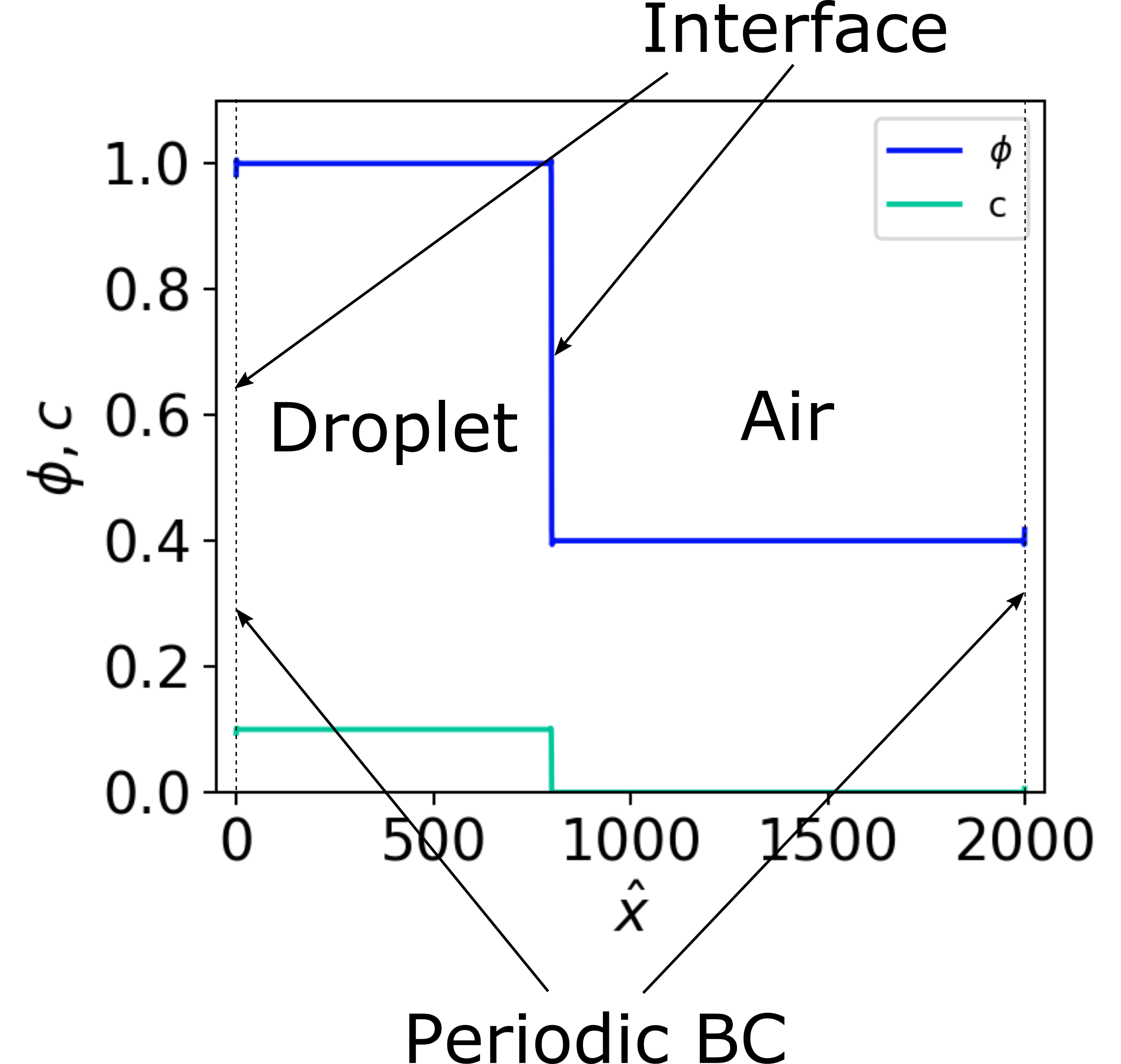}
\caption{\label{fig:s4} Initialized one dimensional system, where $\phi$ and $c$ are piecewise continuous over the simulation domain. 
The droplet and the surrounding air phases are separated by an 
interface, which is smooth during the dynamics. Here $2/5$ of the system is droplet and the remainder is air.} 
\end{figure}

\section{Analytical model}
\subsection{Set-up of the Analytical System}
\label{sec:the_system}
For the binary polymer-water system we create two fields, one (the phase field) for the total mass of the droplet $\phi$, and one for the polymer concentration $c$, which is essentially non-zero only inside the droplets. The two fields are discretized over the same 1D lattice, so that each lattice contains some finite value of $\phi$ and $c$. 

We initialize a system with two phases, the droplet and the surrounding air, separated by an interface, see for example 
\Fig{fig:s4}. In phase 1, inside the droplet, we initialize $\phi = \phi_1$ ($=1$) and $c$ to some other finite value $c=c_0$, where $c_0<\phi_1$. In phase 2, the surrounding air, we initialize $c=0$ and $\phi=\phi_0$ with $\phi_0<\phi_1$, which then qualitatively represents the Relative Humidity (RH). 

Since the amount of solvent in the air phase should not significantly increase during a simulation, as in reality would also not happen to the environment during evaporation, we can safely allow $\phi$ not to be conserved in the surrounding phase. 


Furthermore, we used periodic boundary conditions, which means in the case of the one-dimensional droplet there are two interfaces between the droplet and the air.

\subsection{Applying Cahn-Hilliard to Evaporation of a Binary Polymeric Mixture} 
\begin{figure}[h!]
\centering
\includegraphics[width=0.4\textwidth]{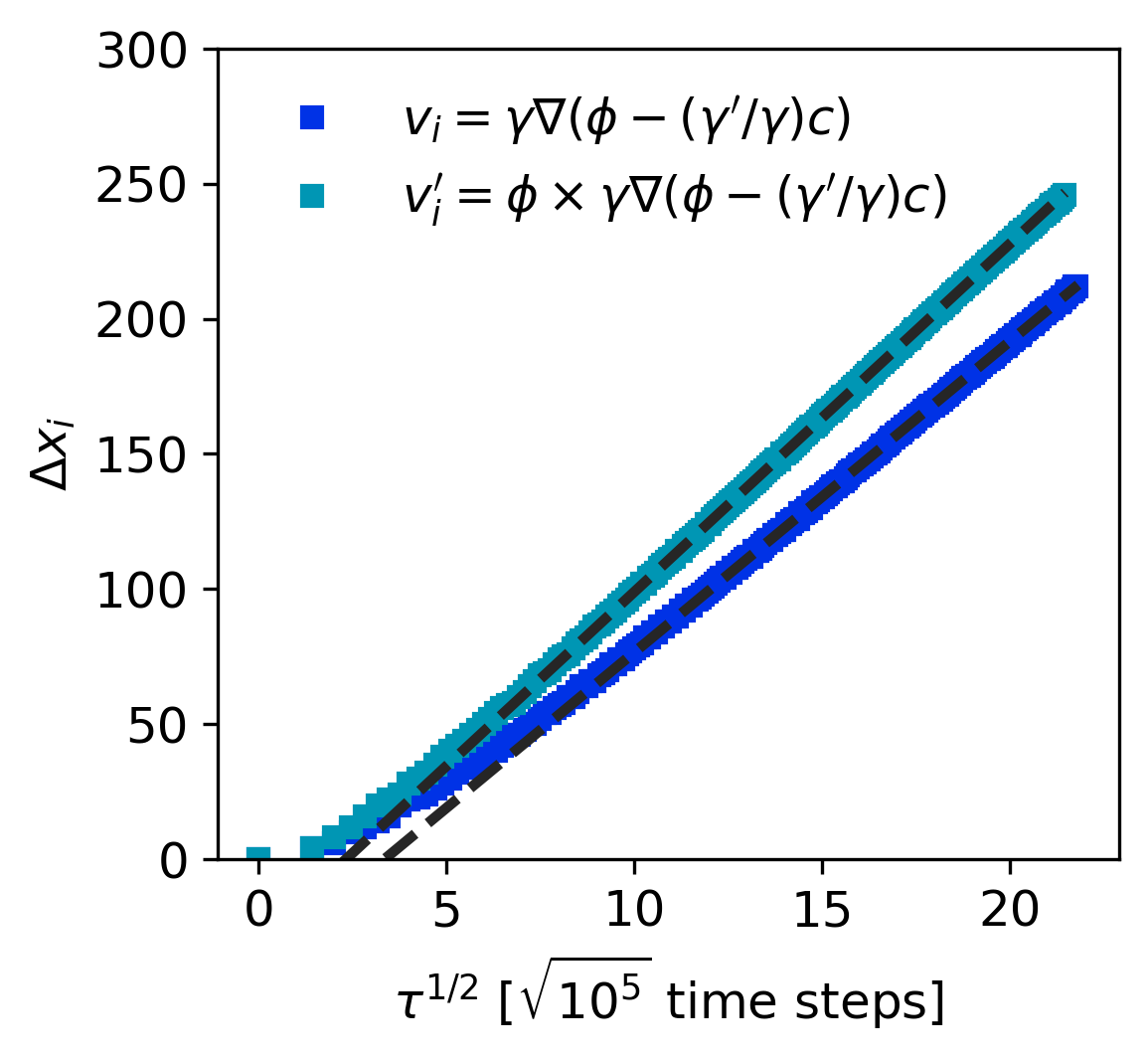}
\caption{\label{fig:s9}Comparison of mass loss $\Delta x_i(t)$ vs $t^{1/2}$ for the evaporation rate expressions $v_i = \gamma \nabla (\phi - \frac{\gamma'}{\gamma}c)$ and $v_i' = \phi \times \gamma \nabla (\phi - \frac{\gamma'}{\gamma}c)$. Both simulations use the exact same parameters for the free energy density $f(\phi,c)$, with $\phi_0=0.4$ and $\gamma^\prime /\gamma=1.5$. Striped black lines are linear fits to $t^{1/2}$.} 
\end{figure}

We start from the general Cahn-Hilliard expression for $\phi$ and add a convective term to account for the advective flux driven by evaporation, similar to \cite{Lee2021}. We consider the following equation for the phase field 
$\phi$:
\begin{align}
	\frac{\partial \phi(x,t)}{\partial t} + v_i \nabla \cdot \phi =  M_\phi \nabla \cdot \left[\nabla \frac{\delta F}{\delta \phi(x)} \right]\,.
\end{align}
In this form, the phase field is not conserved as its time derivative does not equal a total divergence. 
This is due to the $v_i \cdot \nabla \phi$ term. We express the interfacial velocity as $v_i = \gamma \nabla (\phi - \frac{\gamma'}{\gamma}c)$, where $\gamma'$ and $\gamma$ are kinetic parameters. This phenomenological form is chosen because the driving force for evaporation -- hence for the motion of the gas/droplet boundary -- is provided by gradients in the water activity (in our model phenomenologically represented by $\nabla \phi$), which decreases with polymer concentration $c$. 
The ratio $\frac{\gamma'}{\gamma}$ gives the extent to which the polymer reduces the evaporation rate. Using an alternative form for the evaporation rate, that for instance depends on the local value of $\phi$ as $v_i' = \phi \times \gamma \nabla (\phi - \frac{\gamma'}{\gamma}c)$, gives nearly identical results for the interface shrinkage $\Delta x_i$, settling in Diffusion-Limited Evaporation (DLE), \Fig{fig:s9}. Finally, we use the fact that $\frac{\delta F}{\delta \phi} = \mu_{\phi}$ and assume that the mobility $M_{\phi}$ is constant. This leads to the following governing equation for the phase field $\phi$:
\begin{align}
	& \frac{\partial \phi(x,t)}{\partial t} + v_i\cdot \nabla \phi  =  M_\phi \nabla^2 \mu_\phi\,.  \label{eq:phase_field}
\end{align}

We also require an equation for the polymer field, which reads:
\begin{align}
    \label{eq:concentration_field}
	\frac{\partial c(x,t)}{\partial t} + \nabla \cdot (v c) = \nabla \cdot (M_c (c) \nabla \mu_c)\,,
\end{align}
where $c$ is conserved globally and we assume that the mobility of the polymer is dependent on $c$ as $M(c) = \frac{M_0}{1+\beta c}$. The polymer experiences a velocity $v=-v_i$, which is the water velocity that advects the polymer towards the interface.

Note that, another possible interpretation could be of $\phi$ and $c$ as solvent and polymer concentration, respectively. In that case the system would effectively be compressible where there is an increase of $c$ inside the drop. Since the total concentration of material (polymer and solvent) is equal to $\phi+Ac$, where $A$ is a proportionality constant which is not strictly defined, compressibility can be made negligible for small $A$. 

The form of the solvent velocity $v$, and equivalently of the interfacial velocity $v_i$, can also be derived by noting that the force driving solvent flow in the Navier-Stokes equation is $-\nabla p$, with $p$ the pressure of water, which decreases with polymer concentration. As we only use a 1d effective theory, a detailed derivation of the flow field is outside the scope of the current work.

\subsection{Free Energy Density and Chemical Potentials}
To obtain the diffusive term we calculate the chemical potentials as functional derivatives of the free energy density, that depends on the local values of $\phi$ and $c$:
\begin{subequations}
\begin{align}
  \mu_\phi & = \frac{\delta F}{\delta \phi} 
  = \frac{\partial f}{\partial \phi} - \nabla \cdot \frac{\partial f}{\partial \nabla \phi}\,, \label{eq:chem_pot_solvent} \\
  \mu_c & = \frac{\delta F}{\delta c} 
  = \frac{\partial f}{\partial c} - \nabla \cdot \frac{\partial f}{\partial \nabla c}\,. \label{eq:chem_pot_polymer}
\end{align}
\end{subequations}

We start from the following functional form for $f$: 
\begin{equation}
    \label{eq:free_energy_density_final}
    f = \frac{a_1}{4}(\phi - \phi_0)^2(\phi - \phi_1)^2 + \frac{\kappa_\phi}{2}(\nabla \phi)^2 + \frac{\kappa_c}{2}(\nabla c)^2 
    - \frac{a_0}{2}\phi^2 c^2 + \frac{b_0}{2}c^2 + \frac{b_1}{4}c^4 
\end{equation}
Here, the first term $\frac{a_1}{4}(\phi - \phi_0)^2(\phi - \phi_1)^2$ drives phase separation into two phases of compositions $\phi_1$ and $\phi_0$\,; the second and third terms  $\frac{\kappa_\phi}{2}(\nabla \phi)^2$ and $\frac{\kappa_c}{2}(\nabla c)^2$ give the bare surface tension of $\phi$ and $c$\,; the fourth term $\frac{a_0}{2}\phi^2 c^2$ is a phase field phenomenological term which favours the internalisation of the polymer inside the droplet; 
the fifth term $\frac{b_0}{2}c^2$ represents the virial coefficient for polymer diffusion; and the final term $\frac{b_1}{4}c^4$ is the excluded volume interaction of the polymer. 

\subsection{Energy Penalty for Polymer Evaporation}
\label{sec:evaporation_barrier}
The phase field model as outlined in the above is purely interaction based through the free energy density equation. Since polymer and solvent have affinity, in the current form of \Eq{eq:free_energy_density_final} polymer will `leak' into the outer phase. This is not physically unrealistic in liquid-liquid systems, but will become an issue when there is a phase change over the interface like in a gas-liquid system. 

In a gas-liquid system, solvent molecules are easily transferred to the gas phase, \eg\ this does not require much energy. Thus, the energy contribution for $\phi$ to cross the interface and move from liquid to gas can be safely ignored in \Eq{eq:free_energy_density_final}.  

The solute is involatile and unlikely to evaporate, so there should be an energy penalty for $c$ to cross the interface to the outer phase $\phi_0$. This can be implemented by adding a term $g(x)\frac{a_{\text{2}}}{2}c^2$ to our free energy density so that it becomes
\begin{equation}
\label{eq:leaking_barrier}
  f(\phi,c) 
  = \frac{a_1}{4}(\phi - \phi_0)^2(\phi - \phi_1)^2
  + \frac{\kappa_\phi}{2}|\nabla \phi|^2 + \frac{\kappa_c}{2}|\nabla c|^2 
    - \frac{a_0}{2}\phi^2 c^2 + g(x)\frac{a_{\text{2}}}{2}c^2 + \frac{b_0}{2}c^2 + \frac{b_1}{4}c^4\,,
\end{equation}
%
where $g(x)=\Theta(\phi(x)-\frac{\phi_1 - \phi_0}{2})$ is an indicator function which is zero if $\frac{\phi_1 - \phi_0}{2}>\phi(x)$ and one otherwise, 
and $a_{\rm{2}}$ is the energy penalty (units $E/L^2$) for the polymer to be in the outer phase.

\subsection{Concentration Cap}
\label{sec:concentration_cap}
When $c$ increases to some concentration $c=c_g$, the polymer gels and contributes a permanent elastic stress to the system \cite{Leibler1993, Okuzono2008}. This effect can be implemented into the free energy by adding a term $G(x)\frac{K_{\text{g}}}{2}(c(x)-c_{\text{g}})^{2}$, which now becomes
\begin{equation}
\label{eq:concentration_cap}
  f(\phi,c) 
  = \frac{a_1}{4}(\phi - \phi_0)^2(\phi - \phi_1)^2
  + \frac{\kappa_\phi}{2}|\nabla \phi|^2 + \frac{\kappa_c}{2}|\nabla c|^2 - \frac{a_0}{2}\phi^2 c^2 
    + g(x)\frac{a_{\text{2}}}{2}c^2 +  G(x)\frac{K_{\text{g}}}{2}(c-c_{\text{g}})^{2} + \frac{b_0}{2}c^2 + \frac{b_1}{4}c^4 \,,
\end{equation}
where $G(x)=\Theta(c(x)-c_{\rm{g}})$ is another indicator function which is zero if $c(x)<c_{\rm{g}}$ and one otherwise, 
and $K_{\rm{g}}$ (units $E/L^2$)  is the osmotic bulk modulus in the gel phase. 

\subsection{Chemical Potential Expressions}
The chemical potentials used in the diffusive term read as follows:
\begin{subequations}
\label{eq:chem_pot}
\begin{align}
    \mu_\phi  &=  \frac{\partial f}{\partial \phi} - \nabla \cdot \frac{\partial f}{\partial \nabla \phi} =  a_1 (\phi - \phi_0)(\phi - \phi_1)(\phi - \frac{\phi_0 + \phi_1}{2}) - a_0 c^2 \phi   - \kappa_\phi \nabla^2\phi\,,  \label{eq:chem_pot_phi} \\[6pt]    
    \mu_c & = \frac{\partial f}{\partial c} - \nabla \cdot \frac{\partial f}{\partial \nabla c} = -a_0 c \phi^2 + b_0 c + g(x) a_{\rm{2}} c + b_1 c^3 + G(x)K_{\rm{g}}(c-c_{\rm{g}}) -\left[\kappa_c \nabla^2 c \right]\,. \label{eq:chem_pot_c}
\end{align}
\end{subequations}

\newpage

\subsection{Parameter values}

Since the phase field equations are not derived from coarse graining a microscopic model, 
we need to choose all phenomenological parameters such that the system behaves realistically. The set of parameters that was used in our simulations for the free energy and the conservation equations, in units of $E$ (energy $\delta E$), $L$ (length $\delta x$) and T (time $\delta t$), is given in Table~\ref{tab:parameters}. 

\begin{table}[h]
\begin{center}
\caption{Parameter values used in simulations.}
\vspace{5pt}
\label{tab:parameters}
\begin{tabular}{|c|c|c|}
\hline
Parameter     & Units     & Value      \\ \hline
$a_1$         & $E/L^{3}$ & 1.0-3.0    \\ \hline
$\kappa_\phi$ & $E/L$       & 0.1        \\ \hline
$\kappa_c$    & $E/L$       & 0.1-0.4    \\ \hline
$a_0$         & $E/L^{3}$ & 0.02       \\ \hline
$b_0$         & $E/L^{3}$ & 0.001-0.03 \\ \hline
$b_1$         & $E/L^{3}$  & 0.001      \\ \hline \hline
$\gamma$         & $L^{2}/T$  & 0.0001      \\ \hline
$\gamma'$         & $L^{2}/T$  & 0.00005-0.0003      \\ \hline
$M_\phi$         & $L^{3}L^{2}/ET$  & 0.1      \\ \hline \hline
$M_0$         & $L^{3}L^{2}/ET$  & 0.025-0.1      \\ \hline
$\beta$         & - & 0.5-5.0      \\ \hline \hline
$a_{\rm{2}}$         & $E/L^{3}$ & 0.5      \\ \hline 
$K_{\rm{g}}$         & $E/L^{3}$ & 0.5      \\ \hline 
$c_{\rm{g}}$         & - & 0.5-1.0      \\ \hline 
\end{tabular}
\end{center}
\end{table}

\newpage

\subsection{Computational Methods}
\label{sec:4_comp_methods}

To solve the system \Eq{eq:phase_field}, \Eq{eq:concentration_field} and \Eqs{eq:chem_pot}, we used a Lattice-Boltzmann implementation in $C$ that iterates through the following steps:

\begin{enumerate}
    \item Calculate gradients of $\phi$ and $c$ in space using finite difference methods
    \begin{align*}
        \partial f /\partial x &= \frac{f(x+h) - f(x-h)}{2h}\,, \\
        \partial^2 f /\partial x^2 &= \frac{f(x+h) - f(x-h) - 2 f(x)}{h^2}\,; 
    \end{align*}
    \item Calculate $\mu_\phi(x)$ and $\mu_c(x)$ using \Eqs{eq:chem_pot};
    \item Calculate $v_i = \gamma \nabla (\phi - \frac{\gamma'}{\gamma}c)$ and gradients in $\mu_\phi(x)$ and $\mu_c(x)$ using finite difference;
    \item Calculate $\phi_{\text{new}}(x)$ and $c_{\text{new}}(x)$ using \Eq{eq:phase_field} and \Eq{eq:concentration_field}, and update the system.
\end{enumerate}

\newpage

\section{Stability for Initial Conditions}

\begin{figure}[h!]
\centering
\includegraphics[width=\textwidth]{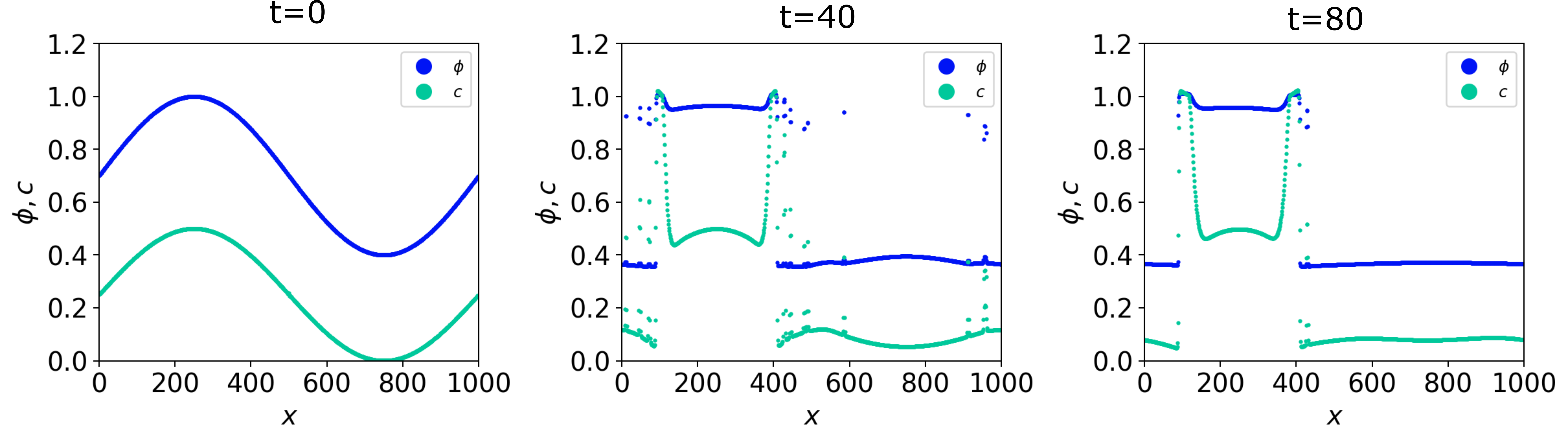}
\caption{\label{fig:s5}Early times evolution of the profiles of $\phi$ and $c$, initialized using sinusoidal initial conditions \Eq{eq:IC_phi} and \Eq{eq:IC_c}.} 
\end{figure}

\begin{figure}[h!]
\centering
\includegraphics[width=0.4\textwidth]{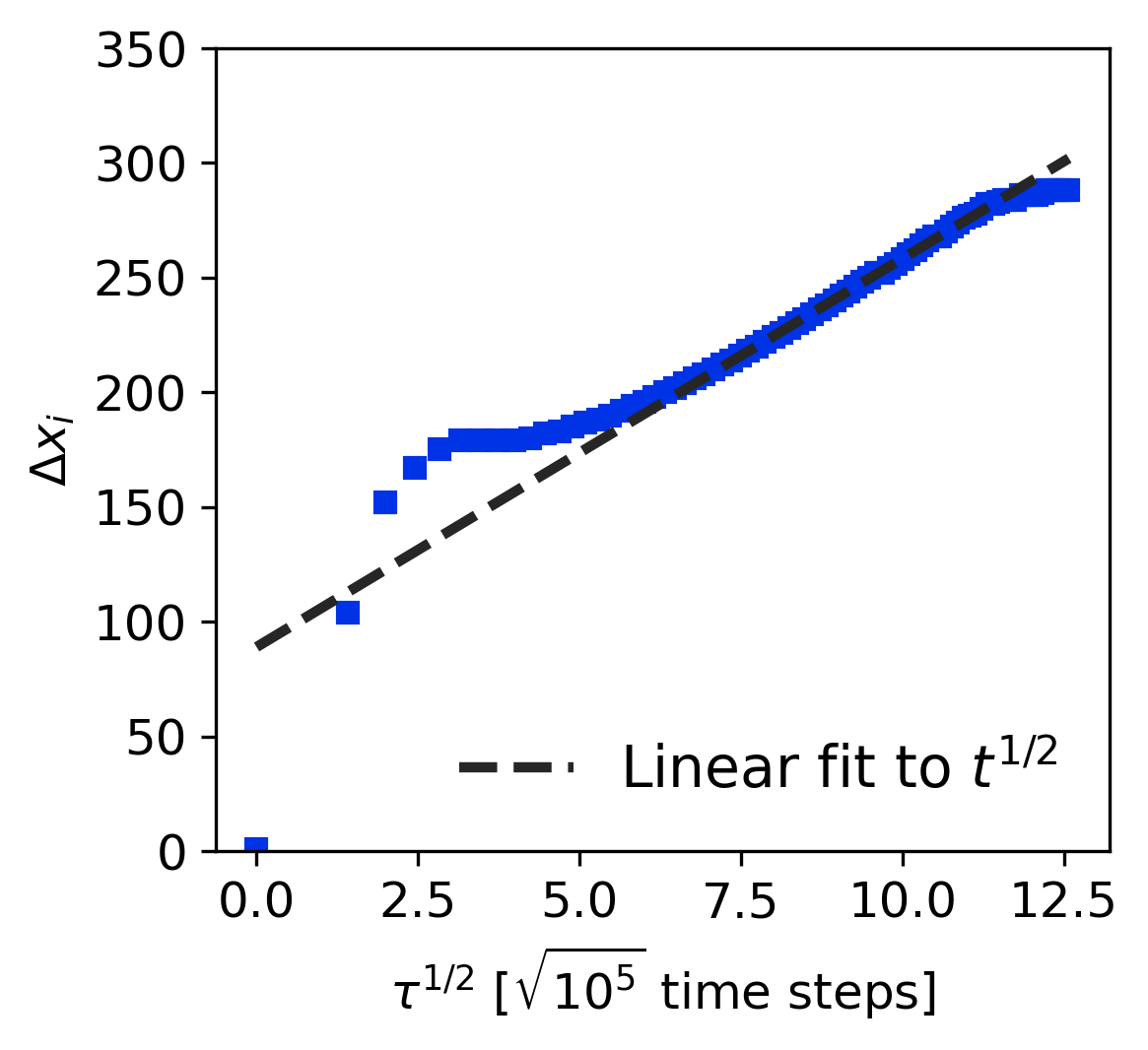}
\caption{\label{fig:s8}Evolution of $m(t) \sim \Delta x_i(t) \sim t^{1/2}$ for system initialized using sinusoidal initial conditions \Eq{eq:IC_phi} and \Eq{eq:IC_c}. Linear fit to $t^{1/2}$ is provided to highlight the settling in DLE.} 
\end{figure}

The functional form of \Eq{eq:concentration_cap} suggests that the system should spontaneously phase separate into a droplet phase and into an environment phase, which entails that the dynamics in the system should be independent of the initial conditions. We test this hypothesis by initializing a one-dimensional system with a sinusoidal distribution of $\phi$ and $c$ as:

\begin{subequations}
\begin{align}
    \phi(x,t=0) & = \frac{\phi_1-\phi_0}{2} \times [\sin{(2 \pi x/L_x)}+1] + \phi_0\,,  \label{eq:IC_phi} \\[6pt]    
    c(x,t=0) & = \frac{c_0}{2} \times [\sin{(2 \pi x/L_x)}+1] \,, \label{eq:IC_c}
\end{align}
\end{subequations}

with $L_x$ the total system size, as shown in \Fig{fig:s5}. Using parameters from the range in Table~\ref{tab:parameters}, after a short initializiation period in which small droplets form in the air phase, \Fig{fig:s5} ($t=40$), we quickly retrieve a similar concentration profile to Fig. 1b in the main manuscript, \Fig{fig:s5} ($t=80$). In the long times regime, the familiar mass loss scaling $m(t) \sim \Delta x_i(t) \sim t^{1/2}$ is observed, \Fig{fig:s8}.

\newpage

\section{Flory-Huggins Implementation in the Model}

\begin{figure}[h!]
\centering
\includegraphics[width=0.4\textwidth]{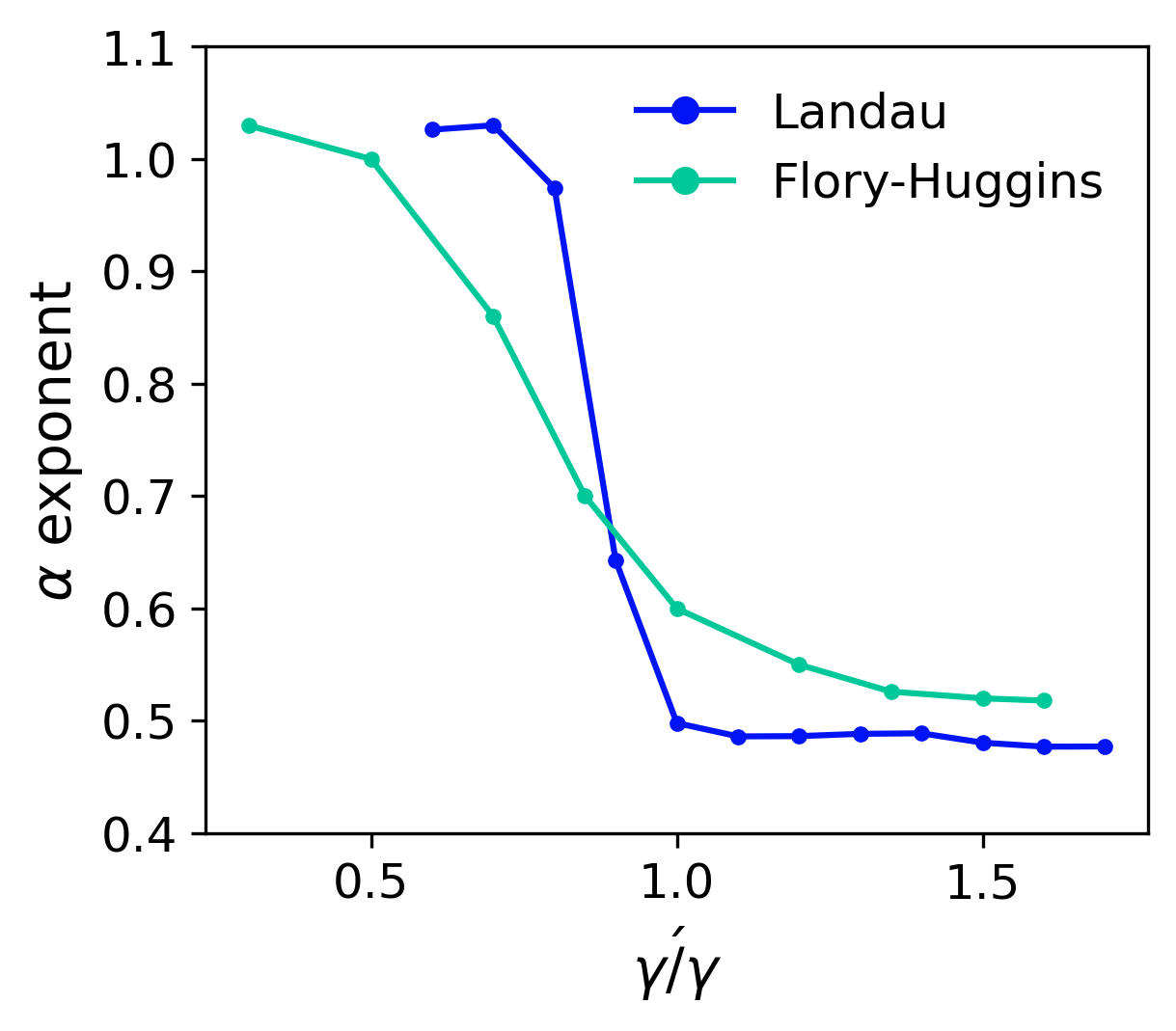}
\caption{\label{fig:s3} Comparison of the evolution of the time exponent $\alpha$ between the Laundau and Flory-Huggins models of the polymeric free energy density, with varying $\gamma'/\gamma$ and $\phi_0=0.40$. Other simulation settings for the Flory-Huggins implementation: $b_0=0.02$ and $\chi=0.001$.} 
\end{figure}

We show how our model can be adapted to using a different polymer model by replacing the Landau expansion for the free energy by a Flory-Huggins model. For large degree of polymerization $N \gg 1$, the free energy density equation becomes:

\begin{equation}
    \label{eq:free_energy_density_final}
    \begin{split}
    f(\phi,c) 
  = \frac{a_1}{4}(\phi - \phi_0)^2(\phi - \phi_1)^2
  + \frac{\kappa_\phi}{2}|\nabla \phi|^2 + \frac{\kappa_c}{2}|\nabla c|^2 - \frac{a_0}{2}\phi^2 c^2 \\[3pt]
    &\hspace{-18em}{} 
    + g(x)\frac{a_{\text{2}}}{2}c^2 +  G(x)\frac{K_{\text{g}}}{2}(c-c_{\text{g}})^{2} + b_0 (1-c)\ln{(1-c)} + \chi c(1-c) \,.
    \end{split}
\end{equation}

Therefore, neglecting any constant contributions, the chemical potential for the polymer now reads:

\begin{equation}
    \label{eq:chem_pot_FH}
    \mu_c = \frac{\partial f}{\partial c} - \nabla \cdot \frac{\partial f}{\partial \nabla c} = -a_0 c \phi^2 - b_0 \ln{(1-c)} + g(x) a_{\rm{2}} c + \chi (1-2c) + G(x)K_{\rm{g}}(c-c_{\rm{g}}) -\left[\kappa_c \nabla^2 c \right]\,.
\end{equation}

The solvent chemical potential remains unchanged.

We compare implementations using a Landau expansion and a Flory-Huggins expression in one line from the state diagrams in Fig 2 of the main manuscript, for varying $\gamma^\prime/\gamma$ and constant $\phi_0=0.4$, in \Fig{fig:s3}. Other simulation settings for the Flory-Huggins implementation are $b_0=0.02$ and $\chi=0.001$. Whilst the behaviour is qualitatively the same, a quantitative difference is observed, where the transition from $m(t)\sim t^1$ to $m(t)\sim t^{1/2}$ is less sharp in using the Flory Huggins equation. We explain this difference through the divergence of $b_0 \ln{(1-c)}$ for $c \to 1$, meaning that there is more spreading near the interface at high $c$ which should lead to slight variations in the settling interface concentration $c_i$, that depend on the driving force, set by $\phi_0$.

\newpage

\section{Evaporation Rate and Growth of Polymer Layer}

\begin{figure}[h!]
\centering
\includegraphics[width=0.50\textwidth]{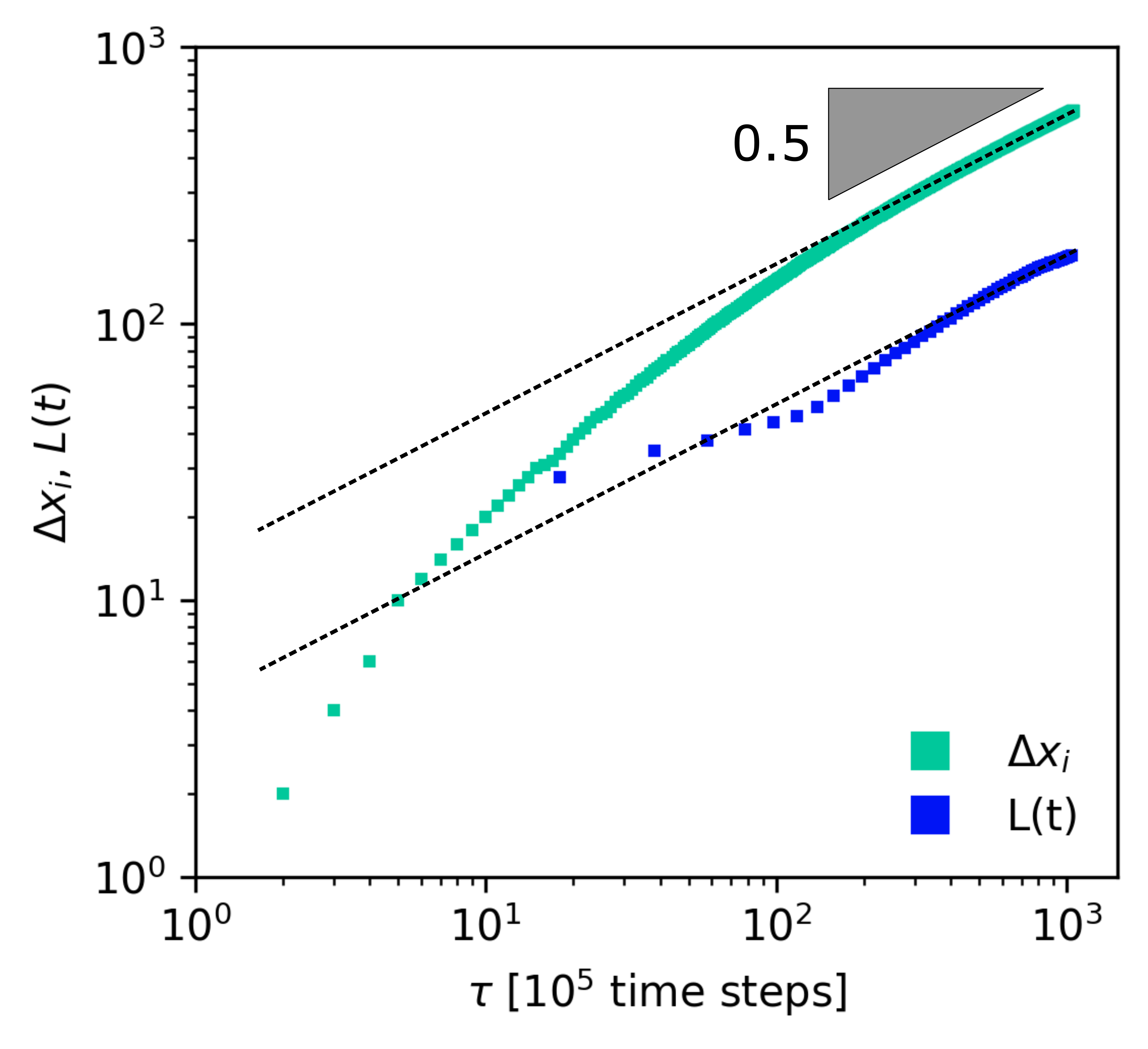}
\caption{\label{fig:s1} Comparison of the evolution of the interface $\Delta x_{\rm{i}}$ and the polymer layer thickness $L(t)$ in a system with $\gamma'/\gamma=1.5$ and $\phi_0=0.35$, plotted logarithmically to highlight the long time power-law behaviour. The capping concentration is reached after approximately $t=100$ timesteps.} 
\end{figure}

%
Our physical interpretation of the DLE regime is that it is due to the buildup of polymer at the interface. 
After the polymer concentration at the interface has reached $c=c_{\rm{g}}$, the polymer layer grows inward as it is swept up by the shrinking interface into the droplet as water molecules diffuse outwards to evaporate. Physically, we therefore expect the scaling law $L(t) \sim m(t)$ to hold, where $L(t)$ is the Full Width Half Maximum (FWHM) of the peak in $c$ near the interface and $m(t)$ is the mass loss. 

To test this scaling law we look at the exponents of the mass loss, e.g. the interface shrinkage $\Delta x_{\rm{i}}$, $m(t) \sim \Delta x_{\rm{i}} \sim t^{\alpha}$ and the polymer layer growth $L(t) \sim t^{\beta}$ in an evaporating droplet where the polymer reaches its capping/gel concentration early in the simulation (Fig. \ref{fig:s1}). We determine the time exponents $\alpha = 0.56$ and $\beta = 0.60$ by power-law fitting, which are indeed similar.